# Regular and chaotic dynamics of nonlinear optomechanical systems controlled by modulated light

A.P. Saiko [1*], G.A. Rusetsky[1], S.A. Markevich[1], R. Fedaruk[2]

[1] Scientific-Practical Material Research Centre, Belarus National Academy of Sciences, Minsk 220072 Belarus
[2] Institute of Molecular Physics, Polish Academy of Sciences, Smoluchowski Str. 17, 60-179 Poznan, Poland

**ABSTRACT**. The nonlinear dynamics of a mechanical resonator in an optomechanical system with linear, quadratic and cubic photon-vibration interactions (with respect to mechanical displacements) in a modulated driving field under conditions of adiabatic elimination of the optical field is studied. Based on the constructed bifurcation diagrams of the mechanical coordinate and the largest Lyapunov exponent as a function of the modulation amplitude, as well as power spectra, phase portraits and Poincaré sections, regions of regular and chaotic dynamics of the optomechanical system are identified. It is also shown that for a certain modulation amplitude in the presence of all three types of interactions, chaotic dynamics of the mechanical resonator (oscillator) is realized, which is replaced by quasi-periodic oscillations in the absence of cubic interaction, and the system returns to chaotic behavior if only linear interaction remains. This non-monotonic dependence of chaotic dynamics on the order of nonlinearity originates from the interplay between parametric driving and effective potential reshaping and manifests that nonlinearity does not always enhance chaos. For an optomechanical system in a membrane-in-the-middle configuration, where only quadratic photon-vibration interaction is present, it is demonstrated that at small modulation amplitudes the mechanical oscillator exhibits quasi-periodic motion in each of the wells of a symmetric two-minimum potential, whereas large modulation amplitudes lead to chaotic motion, involving interwell transitions.

## I. INTRODUCTION

Many real physical dynamical systems consist of several nonlinearly interacting subsystems, leading to their complex motion [1]. Solutions of the corresponding nonlinear dynamical equations exhibit extreme sensitivity to initial conditions and predict phenomena such as attractors [2], period doubling bifurcations, quasiperiodic oscillations, and chaos [3–5]. In optomechanical systems, the nonlinear coupling between the optical and mechanical modes is realized parametrically through radiation pressure, causing oscillations of the mechanical resonator, which in turn affect the behavior of the intracavity optical mode. Such systems are suitable objects both for the study of macroscopic quantum effects [6,7] and for applications of quantum technologies [8]. Optomechanical interactions can lead to cooling [9,10], and amplification of the vibrational modes of a mechanical resonator [11]. In practice, optomechanical systems can be used as sensors to detect small forces [12], displacements [13], masses [14], accelerations [15] with unprecedented accuracy and even lead to the discovery of the quantum nature of gravity [16].

The optomechanical interaction is inherently nonlinear, and generally the evolution of optomechanical systems is described by nonlinear differential equations taking into account the intrinsic high-order interactions, quadratic and cubic in mechanical displacements [17–23]. In recent years, internal optomechanical nonlinear interactions have attracted increasing attention in both quantum [24] and classical [23] domains. In particular, the transition from classical nonlinear dynamics to chaotic dynamics in nonlinear optomechanical systems is investigated, revealing new capabilities for manipulating chaos. Optomechanical chaotic dynamics [25] and optomechanical chaotic oscillation of an on-chip resonator excited by the radiation-pressure nonlinearity [26] have been observed. The multiple time scale dynamics induced by radiation pressure and photothermal effects in a high-finesse optomechanical resonator have been studied experimentally [27]. Chaos-induced stochastic resonance in an optomechanical system and the optomechanically mediated chaos transfer between two optical fields have been demonstrated [28]. Chaotic dynamics in silicon optomechanical oscillators [29] and nanobeams [30] as well as bichromatic synchronized chaos in driven coupled electro-optomechanical nanoresonators [31] have been reported. Recently, chaos has been studied in optomechanical systems coupled to a non-markovian environment [32] and in cavity optomechanical system with Coulomb coupling [33]. Synchronization of chaotic optomechanical system with plasmonic cavity has been proposed for secured quantum communication [34]. Manipulating optical chaos with an electric field in a hybrid electro-optomechanical system has been considered [35]. Chaotic behavior under the influence of a phononic synthetic magnetic field in an optomechanical system has been reported [36]. Currently, the influence of the degree of nonlinearity (up to the third order) of photon-vibrational interaction on the chaos in optomechanical systems is a relevant and insufficiently studied problem. The review [23] emphasizes the absence of studies in which the contribution of higher-order nonlinearities would be considered separately and in combination. At the same time, intrinsic optomechanical nonlinearities play great importance for both understanding fundamental physics and potential applications in sensing with mechanical micro- and nanoresonators [37,38], in data encryption, optomechanical logic and chaos computing as well as in the creation of new types of devices (see [23] and references therein).

*Contact author: saiko@physics.by

In the present paper, unlike previous studies that usually deal with one type of optomechanical nonlinearity, we systematically compare dynamical regimes in optomechanical systems arising from linear, quadratic, and cubic photon-vibration couplings. Throughout we consider the situation in which adiabatic elimination of the cavity optical field is valid. Using modern tools for studying dynamical systems - bifurcation diagrams, Lyapunov exponents, phase portraits, Poincaré sections and power spectra - we identify and characterize regions of periodic motion and chaos in the mechanical resonator's behavior as functions of a certain amplitudes of the control field for given mechanical damping rates, cavity decay rates, and photon-vibration coupling constants. Crucially, we reveal a non-monotonic dependence of chaotic dynamics on the order of nonlinearity: removing cubic coupling can suppress chaos into quasi-periodicity, while retaining only linear interaction restores it. This counterintuitive behavior originates from the interplay between parametric driving and effective potential reshaping, offering a new control knob for chaos engineering in optomechanical sensors and computing devices.

The paper is organized as follows. In Section II, we present a model of the optomechanical system and a nonlinear equation for a mechanical resonator in the adiabatic limit. In Section III, based on numerical solutions of the obtained nonlinear equation, bifurcation diagrams are constructed and an analysis of the dissipative dynamics of the mechanical resonator is performed using phase portraits, Poincaré sections, Lyapunov exponents, and power spectra. Finally, we conclude with a brief summary in Sec. IV. Conditions for the realization of the adiabatic elimination of the optical field variables are discussed in the Appendix.

## II. MODEL HAMILTONIAN AND MAIN EQUATIONS

We study a model optomechanical system containing a mechanical resonator (oscillator) with mass $m$ and frequency $\omega_m$. The nonlinear interaction of this oscillator with the optical mode is described by terms that are linear, quadratic and cubic in the mechanical displacements. The Hamiltonian [18] of this system can be written as

$$H = H_0 + V + V_d \,, \tag{1}$$

$$H_0 = \omega_c \hat{a}^\dagger \hat{a} + \frac{1}{2}\left(\frac{\hat{p}^2}{m} + m\omega_m^2 \hat{x}^2\right), \;\; V = -\hat{a}^\dagger \hat{a} \sum_{n=1,2,3} g_n \hat{x}^n \,,$$

$$V_d = i\varepsilon(\hat{a}^\dagger e^{-i\omega_d t} - H.c.),$$

where $\hat{a}$ ($\hat{a}^\dagger$) is the bosonic annihilation (creation) operators for the cavity mode with $[\hat{a},\hat{a}^\dagger]=1$ (we set the Planck constant $\hbar = 1$), $[\hat{x},\hat{p}] = i$, $\hat{x}$ and $\hat{p}$ are the position and momentum operators for the mechanical oscillator, $\omega_c$ is the cavity frequency, $g_n$ is the optomechanical coupling strength, $n$ = 1,2,3 for linear, quadratic and cubic interactions, respectively, and $V_d$ denotes the optical driving term with the amplitude $\varepsilon$ and frequency $\omega_d$.

From the Heisenberg equations derived from the Hamiltonian (1), we obtain the quantum Langevin equations, in which we make the transition to classical analogies of operators $\hat{a}$ ($\hat{a}^\dagger$), $\hat{x}$ and neglected noise sources [39,40]. Assuming that the decay rate κ of the optical field is much larger than the resonant frequency of the mechanical oscillator, the optical field variables can be adiabatically eliminated from these equations [39]. In this case, for times longer than $\kappa^{-1}$, the equation for the classical momentum $p$ can be obtained in the form:

$$\frac{d}{dt}p = -m\omega_m^2 x + \\ + \frac{\varepsilon^2}{\left[\Delta - \sum_{n=1,2,3} g_n x^n\right]^2 + \kappa^2/4}\frac{d}{dx}\sum_{n=1,2,3} g_n x^n - \frac{\gamma}{2}p, \tag{2}$$

where $\Delta = \omega_c - \omega_d$ is the detuning of the optical field, and $\gamma$ is the phenomenologically introduced mechanical damping rate. It should be noted that at $\kappa \approx \omega_m$, the adiabaticity condition is violated. Conditions for the realization of the adiabatic elimination of the optical field variables are discussed in Appendix. Beyond the adiabatic approximation, the effects of radiation pressure [25,26,41–44] and the dynamical multistability in high-finesse micromechanical optical cavities [45] have been studied.

Thus, in the adiabatic limit, the equation of motion for the mechanical resonator coincides in form with the equation for damped oscillations of a particle with mass $m$ in a nonlinear potential

$$U(x) = m\omega_m^2 x^2/2 + \\ +(2\varepsilon^2/\kappa)\arctan\left[2(\Delta - \sum_{n=1,2,3} g_n x^n)/\kappa\right].$$

In addition to the term representing the initial harmonic potential, this potential contains a term that depends on the amplitude $\varepsilon$ and frequency $\omega_d$ (via detuning $\Delta = \omega_c - \omega_d$) of the driving field, the coupling strengths $g_n$ and the cavity decay rate $\kappa$. Depending on the relationship of the parameters, the potential $U(x)$, as a function of the displacement $x$, can be anharmonic, single-minimum or double-minimum, with symmetrically or asymmetrically located minima.

We assume that the input power of the driving field is modulated with frequency $\Omega$, and $\Omega \ll \kappa$. Therefore, when describing the non-stationary behavior of the mechanical resonator, the input power parameter $\omega_d \varepsilon^2/2\kappa$ changes as [39]

$$\frac{\omega_d}{2\kappa}\varepsilon^2 \to \frac{\omega_d}{2\kappa}(\varepsilon^2 - \varepsilon_M^2 \sin\Omega t),\ \varepsilon_M \le \varepsilon, \qquad (3)$$

where $\varepsilon_M$ is the modulation field amplitude. Next, to express the interaction constants in energy (frequency) units, we make the following substitutions: $g_i \to g_i x_{zpf}^i$, where $x_{zpf} \equiv (1/2m\omega_m)^{1/2}$ is the size of mechanical zero-point fluctuations.

## III. NONLINEAR DYNAMICS OF THE MECHANICAL RESONATOR

### A. Joint linear, quadratic and cubic photon-vibration interactions

In our simulations, we use parameters implemented in modern solid-state optomechanical systems with linear, quadratic and cubic photon-vibration interactions. The mechanical mode frequency $\omega_m$ is chosen to be 10 MHz, the linear coupling strength is $g_1/2\pi = 1$ MHz $(g_1/\omega_m = 0.1)$, the ratio of the quadratic to linear coupling strength is $g_2/g_1 = 0.00075$ $(g_2/\omega_m = 7.5\times10^{-5})$, and the ratio of the cubic to quadratic coupling strength is $g_3/g_2 = 0.00075$ $(g_3/\omega_m = 5.625\times10^{-8})$.

The dynamic system under study is very sensitive to even small changes in its parameters and is characterized by extremely complex motion, which, nevertheless, obeys precise mathematical rules. A powerful tool for studying the complex dynamics of nonlinear systems is the construction of phase portraits and Poincaré sections for them, as well as the determination of the power spectra of oscillations and Lyapunov exponents [40]. As shown in the bifurcation diagram [Fig. 1(a)], the behavior of the mechanical resonator (oscillator) depending on the bifurcation parameter $\varepsilon_M/\varepsilon \equiv \tilde{\varepsilon}$, equal to the ratio of the amplitudes of the modulating and exciting fields, is regular in the range from 0.5 to 0.7. At the chosen parameters of the system, the regular dynamics is a consequence of using the adiabatic limit: $\kappa \gg \omega_m \gg \gamma$. It should be noted that for such a system, beyond the adiabatic approximation $(\kappa/\omega_m = 1)$, alternation of regions of regular dynamics and chaotic behavior of the oscillator was found [17]. The regular behavior of our system persists over a wide range of bifurcation parameter values. After crossing the critical threshold (near $\tilde{\varepsilon} \approx 0.7$), a quasi-periodic and then a chaotic regime is realized. In this case, areas of chaotic behavior can alternate with narrow "windows" of regular dynamics. With a further increase in the bifurcation parameter, the complex dynamics of the system is maintained up to the limit value $\tilde{\varepsilon} = 1$.

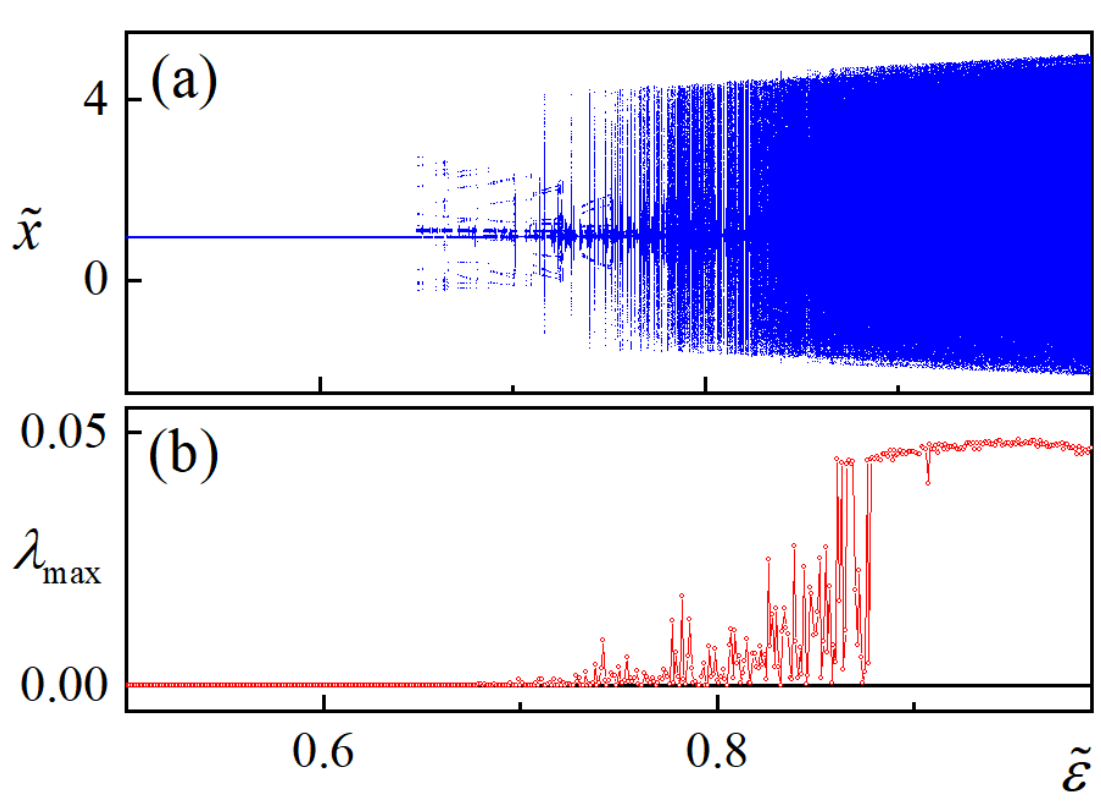

FIG. 1. (a) Bifurcation diagram and (b) the largest Lyapunov exponent for a mechanical resonator (oscillator) of an optomechanical system with linear $(g_1 = 0.1)$, quadratic $(g_2 = 7.5\times10^{-5})$, and cubic $(g_3 = 5.625\times10^{-8})$ photon-vibration interactions. The normalized (with respect to $\omega_m$) parameters are $\Omega = 0.5$, $\gamma = 10^{-3}$, $\kappa = 10$, and $\Delta = -1$. Initial displacement of the mechanical oscillator $x_0 = 0$ and $\tilde{\varepsilon} = \varepsilon_M / \varepsilon$.

For an example, at the bifurcation parameter $\tilde{\varepsilon} = 0.92$, Fig. 2 shows the power spectrum, the phase portrait and the Poincaré sections. The continuous character of the power spectrum, as well as the random distribution of red dots (Poincaré sections) without any noticeable regularity, indicate the chaotic behavior of the system for a given value of the bifurcation parameter. The chaotic behavior of the system is also indicated by the positive value of the largest Lyapunov exponent $\lambda_{\max}$ in this case [Fig. 1(b)].

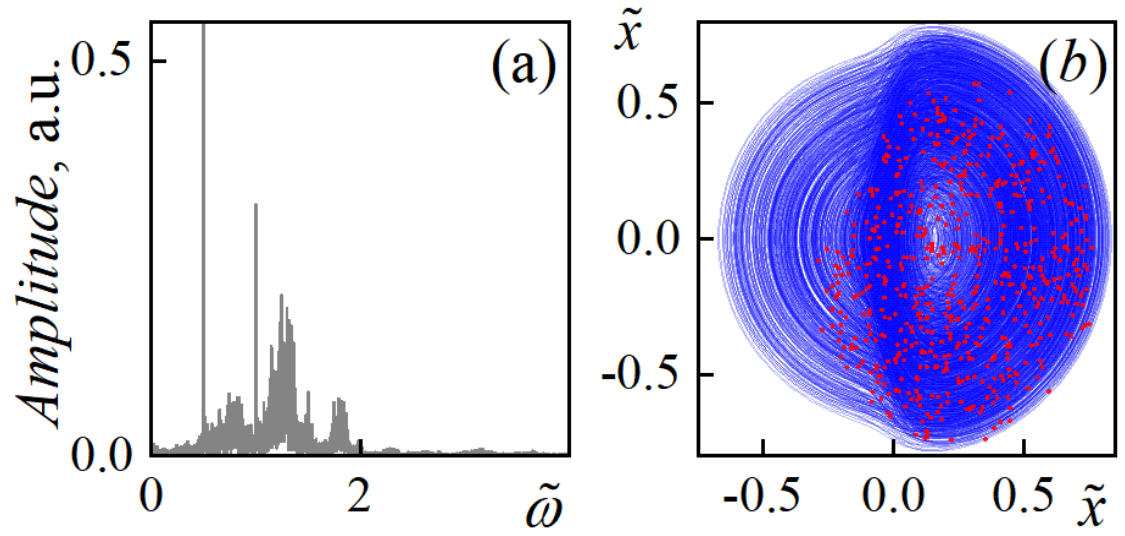

FIG. 2. (a) Power spectrum, (b) phase portrait (blue curves) and Poincaré sections (red dots) for the value $\tilde{\varepsilon} = 0.92$ of the bifurcation parameter. The normalized frequency $\tilde{\omega} \equiv \omega / \omega_m$ is used. Other parameters are the same as in Fig. 1.

When $\tilde{\varepsilon} = 0.7$, the system performs a quasi-periodic motion (Fig. 3) and the power spectrum is represented by the modulation frequency and its harmonics, as well as the frequency of the mechanical oscillator $(\omega_m = 2\Omega)$, somewhat modified due to nonlinear corrections [Fig. 3(a)]. The phase portrait is formed by a band of finite width in the form of an open figure eight, and the dots of the Poincaré map are grouped in the form of a disk (due to small relative phase shifts of oscillations at multiple frequencies) in a narrow region within this band [Fig. 3(b)]; the Lyapunov exponent is near zero [Fig. 1(b)].

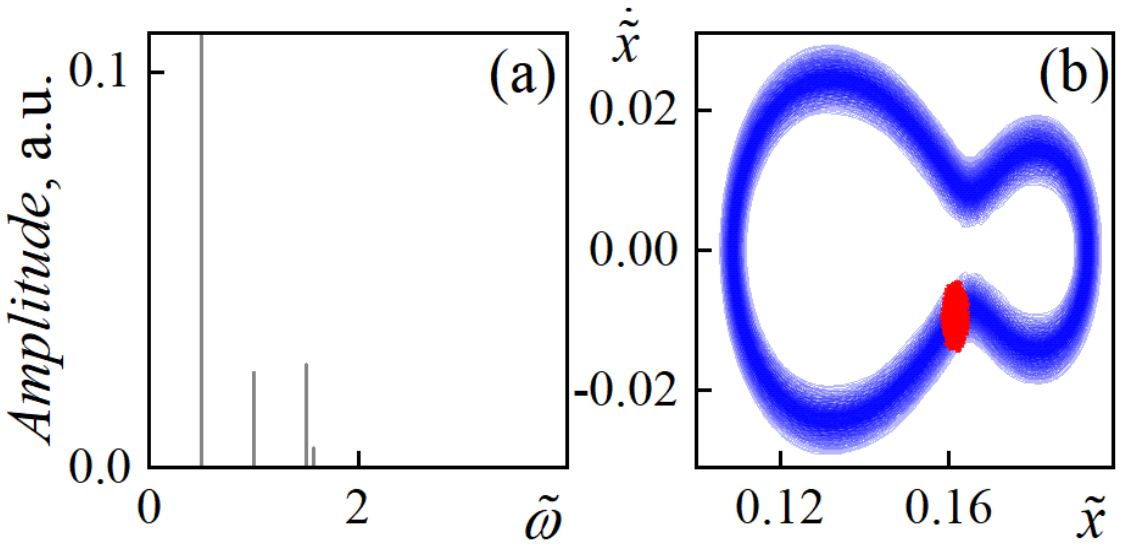

FIG. 3. (a) Power spectrum, (b) phase portrait (blue curves) and Poincaré sections (red dots) for the bifurcation parameter $\tilde{\varepsilon} = 0.7$. Other parameters are the same as in Fig. 1.

It is remarkable that the slightest change in the parameter $\tilde{\varepsilon}$ from 0.7 (Fig. 3)) to 0.709899 (Fig. 4) leads to the system entering a "window" where the periodic motion of the mechanical oscillator is realized at the frequency $\Omega = 0.5$ and its multiples [Fig. 4(a)], and the phase portrait is a closed curve with a single wide (a consequence of computational limitations) Poincaré point [Fig. 4(b)].

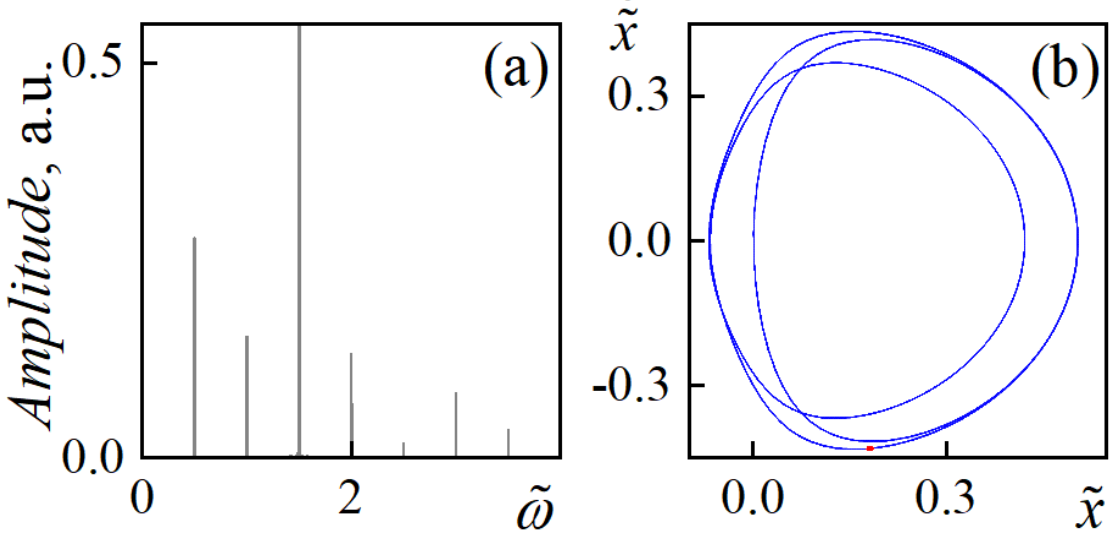

FIG. 4. (a) Power spectrum, (b) phase portrait (blue curves) and Poincaré section (the red dot) for the bifurcation parameter $\tilde{\varepsilon} = 0.709899$. Other parameters are the same as in Fig. 1.

### B. Decreasing degree of nonlinearity photon-vibration interactions

Let us consider the behavior of a mechanical oscillator in an optomechanical system when photon-vibration interactions include (i) linear, quadratic and cubic in mechanical displacements, (ii) linear and quadratic, and (iii) only linear. For all three cases, we choose the same value of the bifurcation parameter, $\tilde{\varepsilon} = 0.829448$. In this case, the change in the behavior of the mechanical oscillator is shown in Fig. 5. In the presence of all three types of interactions $(g_1, g_2, g_3)$, the power spectrum is continuous [Fig. 5(a)], and the Poincaré points on the phase portrait are distributed randomly [Fig. 5(b)], i.e. a chaotic regime is realized.

If the cubic interaction is "switched off" $(g_1 \neq 0,\ g_2 \neq 0,\ g_3 = 0)$, the power spectrum is discrete [Fig. 5(c)] and consists of the modulation frequency, its harmonics, and the frequency of the mechanical oscillator, slightly modified due to nonlinear corrections. This behavior is due to the interaction of parametric control and effective change of potential shape. The phase portrait has the shape of a figure eight with a band of finite width, and the red dots of the Poincaré map form a regular structure in a narrow region of this band [Fig. 5(d)]. In this case the system exhibits quasi-periodic behavior. The oscillator's trajectories are confined only to the positive region of phase space along the axis $x / x_{zpf}$, while for the previous case they extended to the negative half almost symmetrically with respect to the zero line.

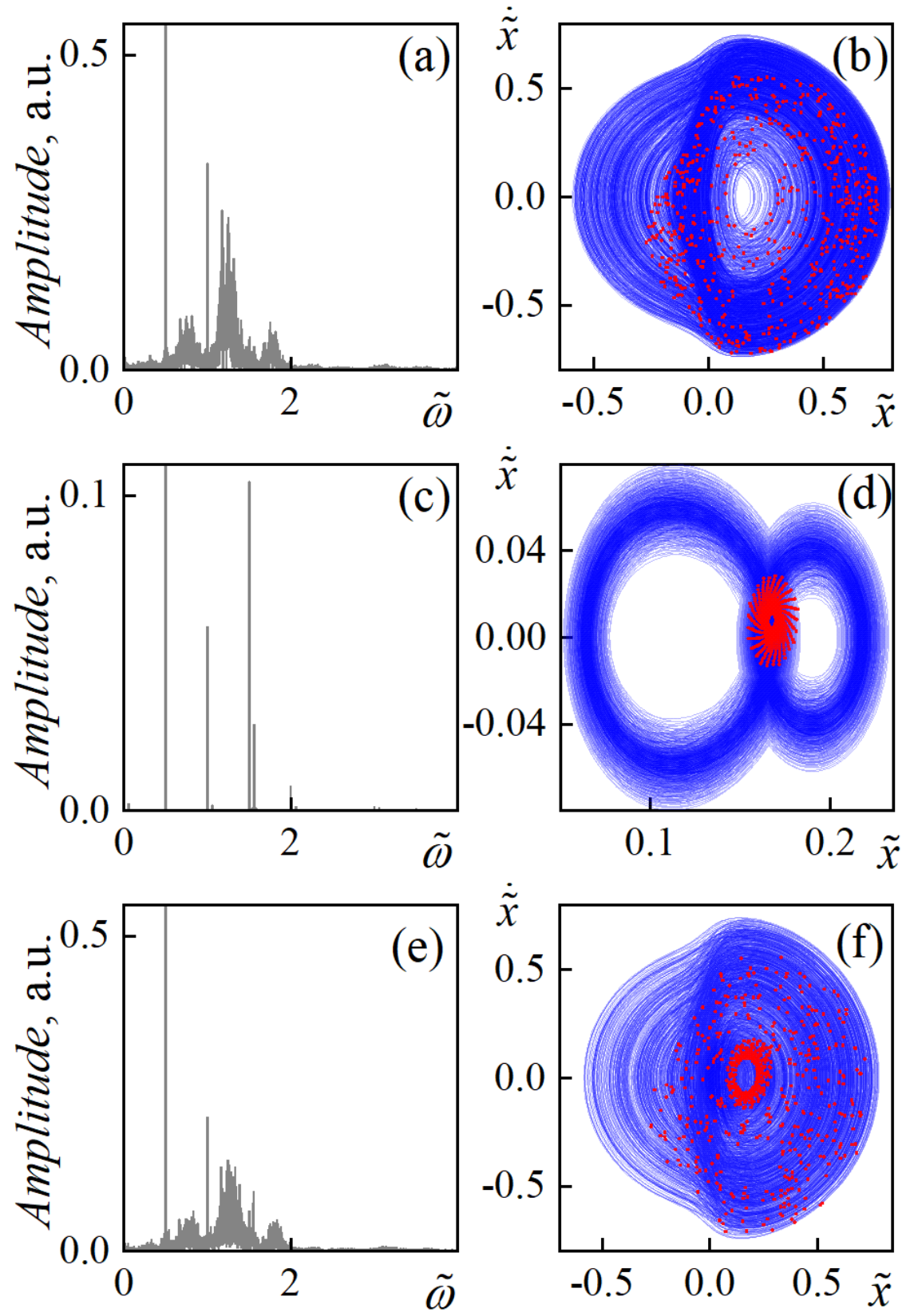

FIG. 5. Power spectra, phase portraits (blue curves) and Poincaré sections (red dots) for the bifurcation parameter $\tilde{\varepsilon} = 0.829448$. (a) and (b) $g_1 = 0.1$, $g_2 = 7.5\times10^{-5}$, $g_3 = 5.625\times10^{-8}$; (c) and (d) $g_1 = 0.1$, $g_2 = 7.5\times10^{-5}$, $g_3 = 0$; (e) and (f) $g_1 = 0.1$, $g_2 = 0$, $g_3 = 0$. Other parameters are the same as in Fig. 1.

The presence of only linear interaction $(g_1 \neq 0,\ g_2 = 0,\ g_3 = 0)$ again leads to chaotic behavior of the oscillator in the negative and positive regions of the phase space [Fig. 5(e), (f)], as in the first case. This non-monotonic transition reveals a subtle interplay between different orders of nonlinearity. The reappearance of chaos in a purely linear regime can be attributed to the action of the modulation field in the presence of the anharmonic effective potential. While cubic nonlinearity introduces amplitude-dependent frequency shifts that promote chaotic mixing through resonance overlap, its removal reveals a quasi-periodic torus stabilized by quadratic effects. However, in the absence of quadratic terms, the system loses this stabilizing effect, and the linear photon-vibration coupling alone is sufficient to drive the oscillator across bifurcation thresholds, restoring chaotic interwell-like transitions. This demonstrates that higher-order nonlinearities do not necessarily enhance chaos; rather, they can suppress it by modifying the effective potential landscape.

### C. A membrane-in-the-middle optomechanical system

Let us consider the dynamics of a mechanical oscillator in an optomechanical system with a membrane-in-the-middle configuration. In this case, the optical mode is tuned to a node of the mechanical displacement, suppressing linear coupling $(g_1 \approx 0)$. The dominant quadratic term $(g_2 \neq 0)$ modifies the effective potential such that, under finite detuning, it develops a symmetric double-well structure, enabling interwell dynamics at high amplitudes of the modulation field. Experimentally realized coupling strengths $g_2 / \omega_m$ ranged from 0.01 to 0.04 at the mechanical frequency of $\omega_m = 355.6$ kHz [22] (see also [46]). As an example, Fig. 6 shows a bifurcation diagram and the largest Lyapunov exponent for a mechanical resonator of such system at $g_2 / \omega_m$ = 0.042. The bifurcation diagram has a two-part structure [Fig. 6(a)]: the behavior of the oscillator in the right part of the cavity (positive values $\tilde{x}$) is almost mirrored in its left part, which corresponds to the motion in a practically symmetric effective double-well potential and is natural for a membrane-in-the-middle optomechanical system. The corresponding behavior of the largest Lyapunov exponent $\lambda_{\max}$ is presented in Fig. 6(b). The largest Lyapunov exponent remains close to zero up to values $\tilde{\varepsilon} \approx 0.5$, and at larger values it fluctuates in the positive region. It can be expected that the dynamics of the system in the range of values $\tilde{\varepsilon}$ from 0 to 0.5 should be quasi-regular and at $\tilde{\varepsilon} > 0.5$ there is a tendency for it to transition to a chaotic regime.

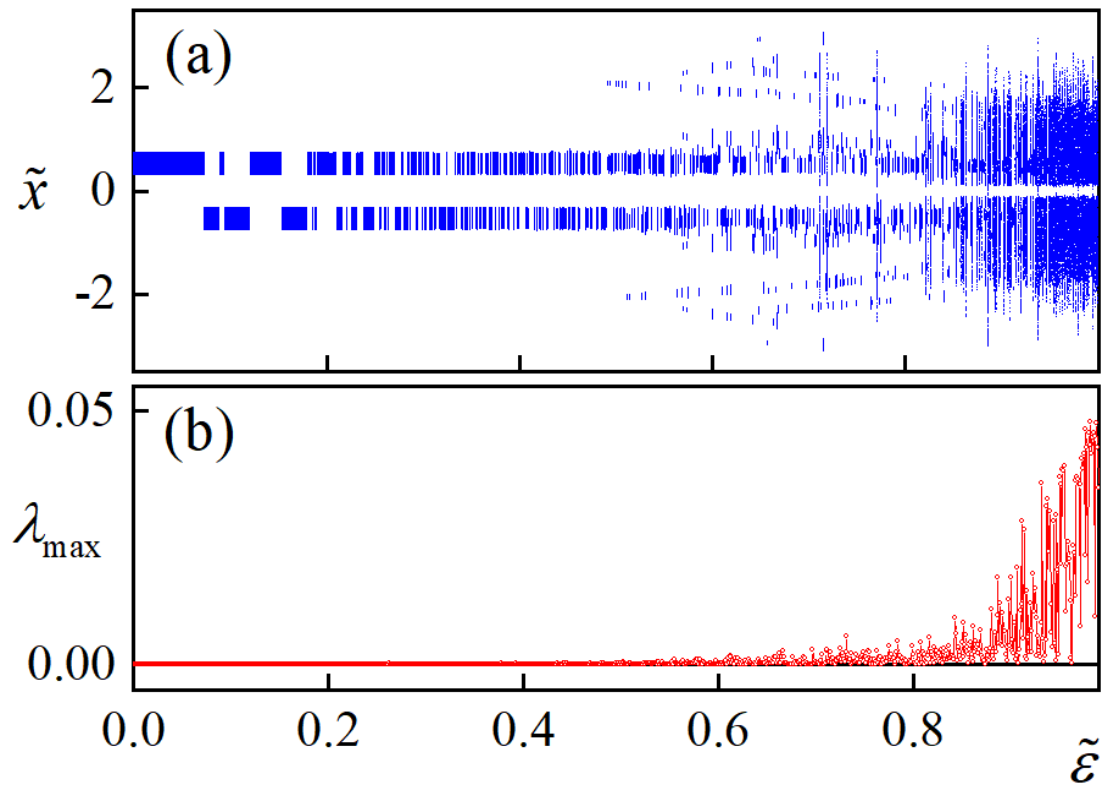

FIG. 6. Bifurcation diagram (a) and the largest Lyapunov exponent (b) for a mechanical resonator of an optomechanical system with a membrane-in-the-middle configuration and quadratic $(g_2 = 0.042)$ photon-vibration interaction. The normalized (with respect to $\omega_m = 355.6$ kHz) parameters are: $\Omega = 1.3$, $\gamma = 10^{-3}$, $\kappa = 10$, $\Delta = -1$. Initial displacement of the mechanical oscillator $x_0 = 0.1$ and $\tilde{\varepsilon} = \varepsilon_M / \varepsilon$.

Figure 7 shows examples of quasi-periodic behavior of the mechanical oscillator using the power spectrum, phase portraits with Poincaré sections for values of the bifurcation parameter $\tilde{\varepsilon}$ equal to 0.4 [Fig. 7(a), (b)] and 0.401 [Fig. 7(c), (d)]. It can be seen that changing this parameter by one thousandth leads to a switch in the oscillator dynamics from the negative region of the phase portrait [Fig. 7(b)] to the positive [Fig. 7(d)].

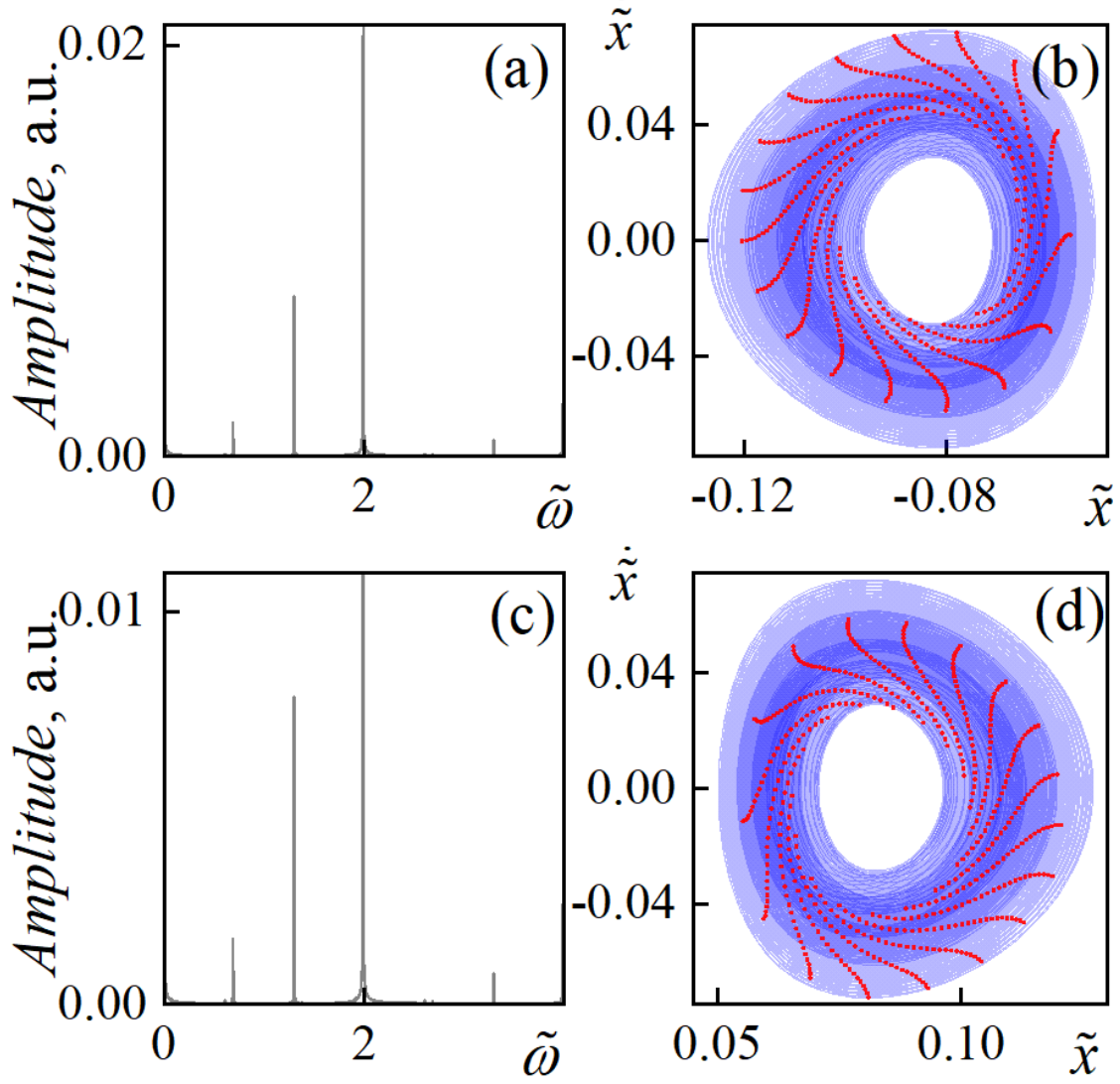

FIG. 7. (a) and (c) Power spectra, (b) and (d) phase portraits (blue curves) and Poincaré sections (red dots) for the two values of the bifurcation parameter: (a) and (b) $\tilde{\varepsilon} = 0.4$; (c) and (d) $\tilde{\varepsilon} = 0.401$. Other parameters are the same as in Fig. 6.

The chaotic regime for $\tilde{\varepsilon} = 0.95$ is shown in Fig. 8. The power spectrum has a (quasi)continuous form [Fig. 8(a)], and on the phase portrait, symmetrically located relative to the zero of the *x*-axis, the points of the Poincaré sections are randomly distributed [Fig. 8(b)]. In this region of large modulation amplitudes, chaotic transitions of the mechanical oscillator between the wells of the two-minimum potential also occur.

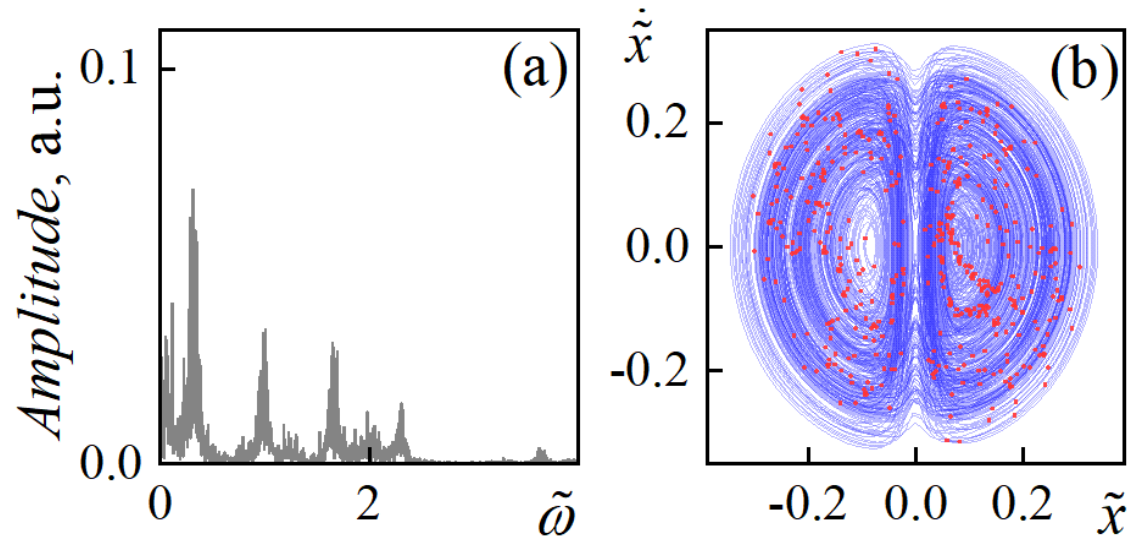

FIG. 8. (a) Power spectrum, (b) phase portrait (blue curves) and Poincaré sections (red dots) for the bifurcation parameter $\tilde{\varepsilon} = 0.95$. Other parameters are the same as in Fig. 6.

## IV. CONCLUSIONS

Using modern tools for studying dynamical systems - bifurcation diagrams, phase portraits, Poincaré sections, Lyapunov exponents, and power spectra - we have studied the dissipative nonlinear dynamics of a mechanical resonator (oscillator) in optomechanical systems with photon-vibration interactions linear, quadratic and cubic in mechanical displacements. Under condition of adiabatic elimination of the optical field, ranges of bifurcation parameter (the ratio of the amplitudes of the modulating and exciting fields) were found, where regular or chaotic dynamics of the optomechanical system are realized. Narrow windows of regular dynamics within broad regions of chaotic behavior were found. At some fixed amplitude of the modulation field, the influence of the degree of nonlinearity of photon-vibration interactions on the behavior of the mechanical resonator was systematically analyzed. A non-monotonic transition between chaos and quasi-periodicity was identified demonstrating that removing cubic nonlinearity can suppress chaos, whereas retaining only linear interaction restores it. Complex dynamics of the mechanical resonator in a membrane-in-the-middle optomechanical system was also considered, where only quadratic photon-vibration interaction is present. A clear separation of intrawell quasi-periodicity and interwell chaos was revealed for this system. It was shown that at small modulation amplitudes the mechanical oscillator exhibits the quasi-periodic motion in a symmetric two-minimum potential, whereas at large modulation amplitudes this motion becomes chaotic involving interwell transitions.

We expect that the high sensitivity of the mixed regular-chaotic dynamics to small parameter changes suggests that such systems can serve as criticality-enhanced sensors. By biasing the system to operate at the "edge of chaos" (near the bifurcation threshold at $\tilde{\varepsilon} \approx 0.7$), minute external perturbations – such as the adsorption of a single molecule (mass sensing) or a femtonewton external force – will induce a macroscopic qualitative change in the oscillator's dynamics, providing a high-contrast signal readout. Moreover, the chaotic interwell transitions observed in the membrane-in-the-middle configuration provide a physical mechanism for True Random Number Generators (TRNGs) for cryptographic applications.

## APPENDIX

In Appendix we show that the calculations presented above satisfy the conditions for implementing the adiabatic elimination of optical field variables. The solution for the time behavior of the optical field $a$, following from the classical treatment of the optomechanical system, will asymptotically approach the solution in the adiabatic approximation for $\kappa \gg \omega_m \gg \gamma$. As a justification for this statement, let's consider the situation when a mechanical oscillator performs approximately sinusoidal oscillations with an unperturbed frequency $\omega_m$: $x = \bar{x} + Ae^{i\omega_m t}$, $\bar{x}$ is the average oscillator position, and A is the oscillation amplitude. Then the expression for the optical amplitude takes the following form [45]:

$$\frac{|a(t)|}{|a_L|} = \left| \sum_{n=-\infty}^{\infty} J_n\left(-\frac{Ag}{x_{zpf}\,\omega_m}\right) e^{in\omega_m t} \Big/ \left(in\omega_m + \frac{\kappa}{2} - ig\frac{\bar{x}}{x_{zpf}}\right) \right|, \tag{A1}$$

where $J_n$ is the Bessel function. A comparison of the solution described by Eq. (A1) and the solution in the adiabatic approximation is shown in Fig. 9. It is clear that the exact solution does not agree with the approximate one at $\kappa = 1$ [45], and in this case it is impossible to adiabatically eliminate the variable $a$ from the classical differential equations for $a$ and $x$. The situation changes with increasing decay rate $\kappa$ of the optical field, and already at $\kappa > 6$ the solution for the amplitude is described fairly well in the adiabatic approximation. Thus, if the decay rate $\kappa$ of the optical field is much larger than the resonant frequency $\omega_m$ of the mechanical oscillator (and also $\kappa \gg \gamma$), the optical field variables can be adiabatically eliminated from consideration, and we obtain equation (2) to describe the dynamics of a mechanical resonator.

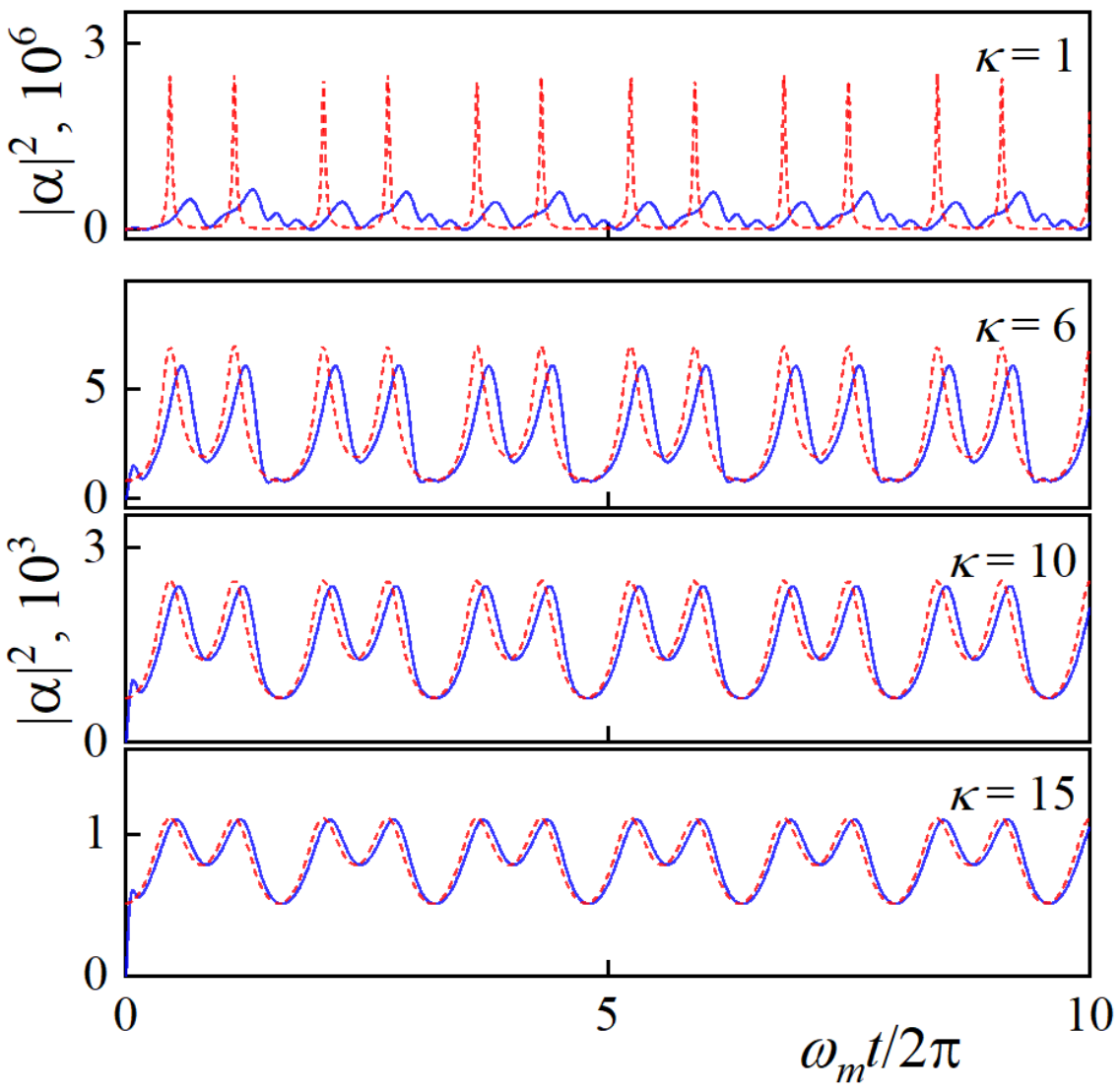

FIG. 9. Time dependence of the optical field intensity for different values of the decay rate $\kappa$. The exact solution and are the result of the adiabatic approximation are shown by solid and dashed lines, respectively. All parameters are normalized with respect to $\omega_m$: $\bar{x}/x_{zpf} = 16.4$, $A/x_{zpf} = 65.4$, $\varepsilon/\omega_m = 250$, $g/\omega_m = g_1 = 0.1$, $g_2 = 0.0$, $g_3 = 0.0$, $\varepsilon \equiv |a_L|$.